\documentclass[%
reprint,
amsmath,
amssymb,
aps,
prd,
]{revtex4-2}

\usepackage{booktabs}
\usepackage{graphicx}% Include figure files
\usepackage{xcolor}% Include figure files
\usepackage{dcolumn}% Align table columns on decimal point
\usepackage{bm}% bold math
\usepackage{subfig}
\usepackage{ragged2e}% Provides justification commands
\usepackage{comment}

\usepackage{ulem}
\newcommand\hlc[2]{\bgroup\markoverwith
  {\textcolor{#1}{\makebox[0.75pt]{\rule[-.5ex]{1pt}{2.5ex}}}}\ULon{#2}}
\definecolor{palatinateblue}{rgb}{0.15, 0.23, 0.89}
\definecolor{Periwinkle}{RGB}{204, 204, 255}

\begin{document}

\preprint{APS/123-QED}

\title{Machine Learning-driven Conservative-to-Primitive Conversion in \\ Hybrid Piecewise Polytropic and Tabulated Equations of State}% Force line breaks with \\

\author{Semih Kacmaz}
\affiliation{Department of Physics, University of Illinois Urbana-Champaign, Urbana, Illinois 61801, USA}
\affiliation{National Center for Supercomputing Applications, University of Illinois Urbana-Champaign, Urbana, Illinois 61801, USA}
\author{Roland Haas}
\affiliation{Department of Physics and Astronomy, University of British Columbia, Vancouver, BC V6T 1Z1, Canada}
\affiliation{Department of Physics, University of Illinois Urbana-Champaign, Urbana, Illinois 61801, USA}
\affiliation{National Center for Supercomputing Applications, University of Illinois Urbana-Champaign, Urbana, Illinois 61801, USA}
\author{E. A. Huerta}
\affiliation{Data Science and Learning Division, Argonne National Laboratory, Lemont, Illinois 60439,
USA}
\affiliation{Department of Computer Science, The University of Chicago, Chicago, Illinois 60637, USA}
\affiliation{Department of Physics, University of Illinois Urbana-Champaign, Urbana, Illinois 61801, USA}

\date{\today}% It is always \today, today,

\begin{abstract}
\noindent We present a novel machine learning (ML)-based method to accelerate conservative-to-primitive inversion, focusing on hybrid piecewise polytropic and tabulated equations of state. Traditional root-finding techniques are computationally expensive, particularly for large-scale relativistic hydrodynamics simulations. To address this, we employ feedforward neural networks ({\texttt{NNC2PS}} %MDPI: Please confirm if the special font needs to be retained.
%Author: While this is not a crucial point, the current styling is being used consistently throughout the text. It could be removed, if it has the potential to cause issues.
and \texttt{NNC2PL}), trained in {PyTorch (2.0+)} %MDPI: Please state the software version number.
%Author: Added. This is not a strict requirement and hence the "2.0+".
 and optimized for GPU inference using NVIDIA TensorRT {(8.4.1)}, %Author: TensorRT version added
 achieving significant speedups with minimal accuracy loss. The \texttt{NNC2PS} model achieves \( L_1 \) and \( L_\infty \) errors of \( 4.54 \times 10^{-7} \) and \( 3.44 \times 10^{-6} \), respectively, while the \texttt{NNC2PL} model exhibits even lower error values. TensorRT optimization with mixed-precision deployment substantially accelerates performance compared to traditional root-finding methods. Specifically, the mixed-precision TensorRT engine for \texttt{NNC2PS} achieves inference speeds approximately 400 times faster than a traditional single-threaded CPU implementation for a dataset size of 1,000,000 points. Ideal parallelization across an entire compute node in the Delta supercomputer (dual AMD 64-core 2.45 GHz Milan processors and 8 NVIDIA A100 GPUs with 40 GB HBM2 RAM and NVLink) predicts a 25-fold speedup for TensorRT over an optimally parallelized numerical method when processing \mbox{8 million} data points. Moreover, the ML method exhibits sub-linear scaling with increasing dataset sizes. We release the scientific software developed, enabling further validation and extension of our findings. By exploiting the underlying symmetries within the equation of state, these findings highlight the potential of ML, combined with GPU optimization and model quantization, to accelerate conservative-to-primitive inversion in relativistic hydrodynamics simulations.
\end{abstract}

\maketitle

%\tableofcontents

\section{\label{sec:intro}Introduction}

In numerical relativity, accurately modeling astrophysical 
systems such as neutron star mergers~\cite{Radice_2020,Ciolfi:2017uak,Kiuchi:2024lpx,PhysRevLett.119.231102,Sun:2022vri,Tsokaros:2021pkh,fernandez2019long,foucart2016low,Camilletti:2022jms,Dietrich:2020eud,Agathos:2015uaa,PhysRevLett.111.131101,RevModPhys.89.015007,RevModPhys.80.1455} relies on solving the 
equations of relativistic hydrodynamics, which involve 
the inversion of conservative-to-primitive (C2P) variable 
relations~\cite{2006ApJ...641..626N,faber2012binary,2005PhRvD..72b4028D}. 
This process typically requires computationally 
expensive root-finding algorithms, such as Newton-Raphson 
methods, and interpolation of complex, multi-dimensional
equations of state (EOS) tables~\cite{2000LRR.....3....2F,Chang:2020ktl}. 
These methods, while robust, incur significant computational
costs and can lead to inefficiencies, particularly in large-scale simulations, where up to billions of C2P calls
may be required per time step. The inherent complexity of this mapping, however, often conceals underlying symmetries and lower-dimensional relationships that a machine learning model can be trained to recognize and exploit.

In view of these considerations, and taking into account 
the advent of GPU-based 
exascale supercomputers such as Aurora and Frontier 
and ongoing efforts to port relativistic hydrodynamics software 
into GPUs~\cite{Kalinani:2024rbk,Zhu:2024utz,Liebling:2020jlq}, this work explores the use of machine learning (ML) 
algorithms that leverage GPU-accelerated computing for 
C2P conversion. 
CPU-based algorithms for C2P conversion typically
involve an iterative non-linear root finder, for which the number of
iterations required to achieve a given target accuracy depends on the
input data, resulting in different runtimes for different points of the
numerical grid. This limits the potential to use SIMD (for CPUs) or
SIMT (for GPUs) parallelism, reducing the effective rate of conversion
achievable using these schemes. An ML approach with its more
predictable runtime and regular memory access pattern may help
alleviate these issues.
Indeed, this work is motivated by recent 
studies that have explored 
the potential of ML to replace traditional 
root-finding approaches for C2P inversion~\cite{dieselhorst21}. 
Specifically, neural networks have shown promise in 
accelerating the C2P inversion process while maintaining 
high accuracy~\cite{dieselhorst21}.
Building on this, the present work introduces 
a novel approach that leverages ML to accelerate 
the recovery of primitive variables from conserved variables 
in relativistic hydrodynamics simulations, with particular 
focus on hybrid piecewise polytropic and tabulated EOS. These 
EOS models provide more realistic descriptions of the dense 
interior of neutron stars, yet their complexity makes the 
traditional C2P procedure very computationally expensive.

To help address these computational challenges, we present a suite of feedforward 
neural networks trained to directly map conserved 
variables to primitive variables, bypassing the need for 
traditional iterative solvers. In particular, we employ a 
hybrid approach, utilizing the flexibility of neural networks 
to handle the challenges posed by complex EOS models. Our models 
are implemented using modern deep learning tools, such as PyTorch, 
and optimized for GPU inference with NVIDIA TensorRT ~\cite{ansel2024pytorch}. Through 
comprehensive performance benchmarking, we demonstrate that our 
approach significantly outperforms traditional numerical 
methods in terms of speed, particularly when using 
mixed-precision deployment on modern hardware accelerators 
like NVIDIA A100 GPUs in the Delta supercomputer.

We evaluate the scalability of our ML models by comparing their inference performance against a single-threaded CPU implementation of a traditional numerical method from the RePrimAnd library~\cite{Kastaun2021}. The benchmark was conducted on a Delta supercomputer compute node, featuring dual AMD 64-core 2.45 GHz Milan processors, 8 NVIDIA A100 GPUs (40 GB HBM2 RAM), and NVLink. For dataset sizes ranging from 25,000 to 1,000,000 points, the numerical method exhibited linear scaling of inference time. In contrast, TensorRT-optimized and TorchScript-based neural networks achieved substantially faster inference, typically demonstrating sub-linear scaling. We investigate two feedforward neural network architectures: a smaller network (\texttt{NNC2PS}) and a larger one (\texttt{NNC2PL}). Notably, mixed-precision TensorRT engines delivered impressive performance, with the \texttt{NNC2PS} engine processing 1,000,000 points in 8.54 ms, compared to {3490} %MDPI: Commas are only used for numbers with five or more digits. We have removed them in four-digit numbers. Please confirm.
%Author: I confirm.
 ms for the numerical method. Ideal parallelization across the entire node (64 CPU cores that support up to 128 threads and 8 GPUs) suggests a 25-fold speedup for TensorRT over the optimally parallelized numerical method when processing 8 million points. These results demonstrate the scalability and efficiency of our ML-based methods, offering significant improvements for high-throughput numerical relativistic hydrodynamics simulations.

This article is structured as follows. Section~\ref{sec:met} introduces 
the EOS considered in this study, along with the methodologies employed 
for designing, training, validating, and testing the ML models. In 
Section~\ref{sec:res}, we present our key results, including an 
assessment of the accuracy of the ML models across different model types 
and quantization schemes. Additionally, we provide a comparison of 
the computational performance of the ML models relative to traditional 
root-finding methods. Finally, Section~\ref{sec:end} offers a summary 
of the findings and outlines potential avenues for future research.

\section{\label{sec:met}Methods}

We present an ML-based model with the potential to accelerate the recovery of primitive variables from conserved variables 
in general relativistic hydrodynamics (GRHD) simulations, specifically 
focusing on scenarios employing hybrid piecewise polytropic EOS
and tabulated EOS. As in traditional approaches, this conversion requires inverting the conservative-to-primitive map, a process often reliant on computationally expensive root-finding algorithms. While previous work has demonstrated the success of machine learning for this task with the $\Gamma$-law EOS~\cite{dieselhorst21}, here, we investigate its application to hybrid piecewise polytropic EOS, which offers a more realistic representation of neutron star interiors, as well as the tabulated EOS, which incorporates the current nuclear physics model of neutron matter. To evaluate the performance of our neural network, we use a traditional CPU-based root-finding algorithm (provided by the RePrimAnd library) as a baseline for comparison. Our aim is to demonstrate the speed advantages of the neural network approach for conservative-to-primitive variable conversion. Our network is implemented using {PyTorch (2.0+) }%MDPI: Please state which version of the software was used.
%Author: Fixed.
  and the inference speed tests are performed {using} %MDPI: Please state which version of the software was used.
  %Author: libtorch comes with PyTorch. TensorRT version is added and highlighted.
 {\texttt{libtorch} and NVIDIA TensorRT {(8.4.1)}'s \texttt{C++} API.} While our numerical experiments are conducted in flat spacetime for simplicity, the C2P inversion is a local operation. Therefore, our method is directly applicable to general relativistic hydrodynamics simulations without loss of generality, as one can always perform the inversion in a local inertial frame.

In general relativity, the equations of relativistic hydrodynamics can be expressed in a conservation form suitable for numerical implementation. Specifically, in a flat spacetime, they constitute the following first-order, flux-conservative hyperbolic {system:} %MDPI: Please confirm if the bold formatting in equations are necessary; if not, please remove it. The following highlights are the same. Please ensure all variables/values in the equation appear in the same format in the text (normal/italic/bold/subscript/superscript).
%Author: Yes, bold formatting is necessary.
\vspace{-3pt}

\begin{equation}
    \label{eq:grhd_cons}
    \frac{1}{\sqrt{-g}} \left( \frac{\partial \sqrt{\gamma} \,\mathbf{u}}{\partial x^0} + \frac{\partial \sqrt{-g} \,\mathbf{F}^i (\mathbf{u})}{\partial x^i} \right) = \mathbf{0},
\end{equation}
where $g = \det(g_{\mu\nu})$ is the metric determinant, and $\gamma = \det(\gamma_{ij})$ is the determinant of the three metrics induced on each spacelike hypersurface. The state vector of the conserved variables is $\mathbf{u}=(D, \mathbf{S}_j, \tau)$, and the flux vector is given by \vspace{-3pt}

\begin{equation}
    \mathbf{F}^i = \left( D \left( v^i - \frac{\beta^i}{\alpha} \right), S_j \left( v^i - \frac{\beta^i}{\alpha} \right) + p \delta^i_j, \tau \left( v^i - \frac{\beta^i}{\alpha} \right) + p v^i \right)\,,
\end{equation}
where $\alpha$ is the lapse function and $\beta^i$ the spacelike shift vector: two kinematic variables describing the evolution of spacelike foliations in spacetime as in a typical \mbox{$3+1$ (ADM) formulation}.

The five quantities satisfying Equation (\ref{eq:grhd_cons}), all measured by an Eulerian observer sitting at a spacelike hypersurface, are the relativistic rest-mass density, $D$, the three components of the momentum density, $S_j$, and the energy density relative to the rest mass density, $\tau = E - D$, respectively. These are related to the primitive variables; rest-mass density, $\rho$, three-velocity, $v_i$, specific internal energy, $\epsilon$, and pressure, $p$ through \vspace{-6pt}
\begin{align}
\label{eq:cons}
    D &= \rho W\,, \nonumber\\
    S_j &= \rho h W^2 v_j\,, \\
    \tau &= \rho h W^2 - p - D, \vspace{-6pt}\nonumber   
\end{align} 
where $W=1/\sqrt{1-\gamma_{ij}v^iv_j}$ is the Lorentz factor, and $h = 1 + \epsilon + p/\rho$ is the specific enthalpy.

Incorporating the EOS into the picture provides the thermodynamical information linking the pressure to the fluid's rest-mass density and internal energy, which, combined with the definitions above, closes the system of equations given in Equation (\ref{eq:grhd_cons})~\cite{Valencia1997, marti2003, font2008}. 

We will first focus on the hybrid piecewise polytropic EOS. The hybrid piecewise polytropic EOS was introduced for simplified simulations of stellar collapse to model the stiffening of the nuclear EOS at nuclear density and include thermal pressure during the postbounce phase~\cite{janka1993}. In gravitational-wave science, it is more commonly used as described in Read et al.~\cite{read2009}, where it enables gravitational-wave parameter estimation and waveform modeling by effectively capturing macroscopic neutron star observables with minimal parameters. The structure of this EOS consists of multiple cold polytropes, defined by parameters $K_0, K_1, \cdots, K_{\mathrm{nsegments}-1}$ and $\Gamma_0, \Gamma_1, \cdots, \Gamma_{\mathrm{nsegments}-1}$, where \texttt{nsegments} denotes the total number of segments. Additionally, it includes a thermal $\Gamma-$law component characterized by $\Gamma_\mathrm{th}$. Continuity of pressure and internal energy across segments, in accordance with the first law of thermodynamics, is ensured after appropriately setting initial values for the polytropic indices, density breakpoints (denoted $\rho_{\mathrm{breaks}})$, and other relevant parameters. For this EOS, the polytropic indices ($\Gamma_i$), the density breakpoints ($\rho_{\mathrm{breaks}}$), and the first segment's polytropic constant ($K_0$) are treated as free parameters. Subsequent constants ($K_i$ for $i > 0$ and all $a_i$) are then determined by enforcing continuity of pressure and internal energy across the breakpoints. In this context, pressure and specific internal energy components in each density interval are given by\vspace{-6pt}
\begin{align}
\label{eq:prim}
    p_{\text{cold}} &= K_i \rho^{\Gamma_i}, \nonumber \\
    \epsilon_{\text{cold}} &= a_i + \frac{K_i}{\Gamma_i - 1} \rho^{\Gamma_i - 1}, \\
    p_\text{th} &= (\Gamma_\text{th} - 1) \rho (\epsilon - \epsilon_{\text{cold}}), \nonumber \\
    p &= p_{\text{th}} + p_{\text{cold}}, \nonumber
\end{align}
where $a_i$ is the segment-specific constant, and the rest mass density, $\rho$, is assumed to fall into the segments specified by each of the $\rho_{\text{breaks}}$. These equations apply to  segment $i$, where the rest-mass density $\rho$ is in the range $\rho_{\mathrm{break}, i -1} < \rho < \rho_{\mathrm{break},i}$.

In addition to the hybrid piecewise polytropic EOS-based model, we will train a separate network to infer the conservative-to-primitive transformation utilizing the tabulated EOS data. Specifically, we will use the Lattimer-Swesty EOS with a compressibility parameter $K=220$ (hereafter referred to as \texttt{LS220} EOS), due to its prevalence and historical significance. Our training dataset is based on a modern, updated version of \texttt{LS220} EOS constructed and made available by Schneider, Roberts, and Ott in a more recent study~\cite{SRO2017}.

Below, we outline the dataset preparation, model architecture, training process, and methods used in inference speed testing with \texttt{libtorch} and NVIDIA TensorRT to evaluate computational efficiency. 
\subsection{Data}
\subsubsection{Piecewise Polytropic EOS-Based Model Data\label{sss:piecewisedata}} 
\begin{figure*}[htbp]
\centering
\includegraphics[width=1.0\textwidth, keepaspectratio]{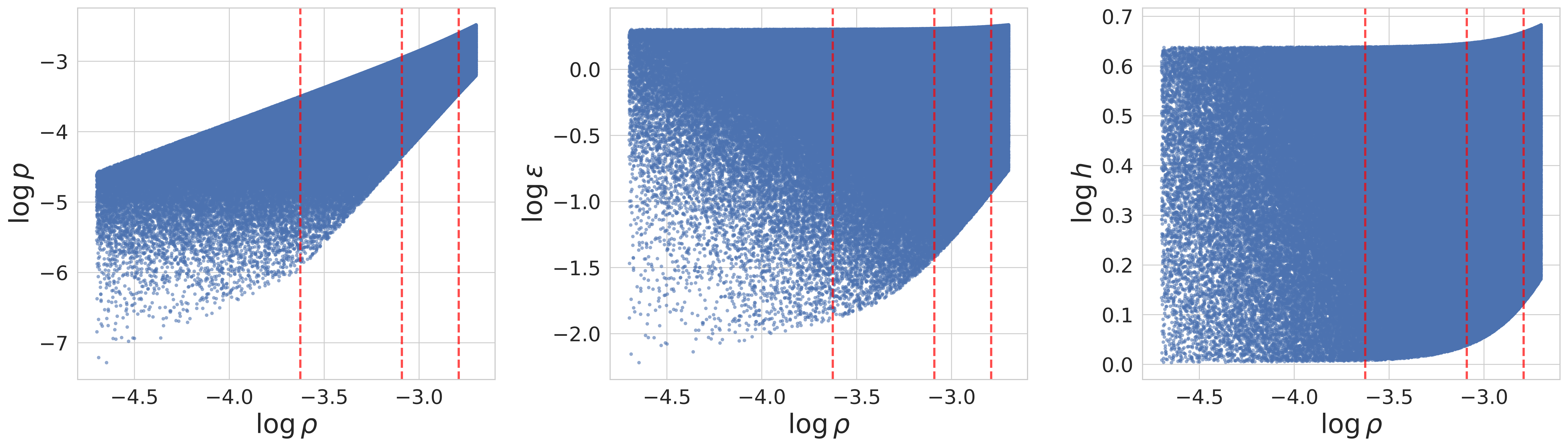}

\caption{\justifying Visualization of the thermodynamic relations based on the complete training data generated for the four-segment piecewise polytropic EOS-based model. From left to right: pressure ($p$) vs. rest-mass density ($\rho$), specific internal energy ($\epsilon$) vs. rest-mass density ($\rho$), and specific enthalpy ($h$) vs. rest-mass density ($\rho$). All quantities are plotted on a logarithmic scale. The distinct segments of the piecewise polytropic EOS are delineated by the red vertical lines.}
\label{fig:hybrid_data}
\end{figure*}

{We generate a dataset of 500,000 samples using geometrized units where $G=c=M_{\odot}=1$}. Without loss of generality, we furthermore use a Minkowski metric $g_{\mu\nu} = \mathrm{diag}(-1,+1,+1,+1)$. The rest-mass density, $\rho$, is sampled uniformly from \mbox{$[2\times 10^{-5}, 2 \times 10^{-3}]$}, and the fluid’s three-velocity is assumed one-dimensional along the $x$-axis, sampled uniformly from $v_x \in (0, 0.721)$. These ranges are chosen to be representative of the conditions found in binary neutron star mergers and to facilitate a direct comparison with the previous work in \cite{dieselhorst21}. Following Ref.~\cite{read2009}, we use an SLy four-segment piecewise polytropic EOS with segment-wise polytropic indices $\Gamma = [1.3569, 3.0050, 2.9880, 2.8510]$. The first segment’s polytropic constant, $K_0$, is set to $8.9493 \times 10^{-2}$. Subsequent polytropic constants, $K_i$, are determined by enforcing pressure continuity. Similarly, the first segment’s constant, $a_0$, is set to zero, while subsequent $a_i$ values ensure continuity of internal energy. The density breaks for the segments are specified at $\rho = 2.3674 \times 10^{-4}$, $8.1147 \times 10^{-4}$, and $1.6191 \times 10^{-3}$. The thermal component has an adiabatic index of $\Gamma_\mathrm{th} = 5/3$. Additionally, the thermal component of the specific internal energy, $\epsilon_\mathrm{th}$, is sampled uniformly from $[0, 2]$ (where $\epsilon_{\mathrm{th}} = \epsilon - \epsilon_{\mathrm{cold}}$). A structured dataset is then constructed by converting the primitive variables to conserved variables using the standard relativistic hydrodynamic relations given in Equation (\ref{eq:cons}). In this dataset, conserved variables serve as input features, and the pressure is the target variable. The resulting dataset is then split into training, validation, and test sets, with each set fully standardized to zero mean and unit variance to ensure equal contribution of all features during neural network training (Figure~\ref{fig:hybrid_data}).\vspace{-3pt}

\subsubsection{Tabulated EOS-Based Model Data}
To generate the training data for the tabulated EOS-based model, we sample from a provided EOS table and follow a procedure similar to the one described in Section~\ref{sss:piecewisedata}. We begin by reading in the EOS table, which contains the variables electron fraction ($Y_e$), temperature ($T$), rest-mass density ($\rho$), specific internal energy ($\epsilon$), and pressure ($p$). These quantities are stored in logarithmic form in the table and are extracted accordingly. For each data point, a random one-dimensional three-velocity, $v_x$, is sampled uniformly on a linear scale from the interval $(0, 0.721)$. Values for electron fraction and temperature are also sampled uniformly on a linear scale from their respective ranges in the table. The rest-mass density is chosen by randomly selecting one of the grid points from the table, which are logarithmically spaced. For this study, we fetched the corresponding values of $p$ and $\epsilon$ directly from the table without interpolation to ensure the training data perfectly represents the tabulated EOS. Using these, the corresponding values of $\rho$, $\epsilon$, and $p$ are then fetched from the EOS table. The primitive variables are then converted into conserved variables using standard relativistic hydrodynamics relations given in Equation (\ref{eq:cons}). A total of 1,000,000 data points are generated using this process ~\cite{Wouters24}. Similarly to the hybrid piecewise polytropic EOS-based model, the data is split into training, validation, and test sets, with each set fully standardized to zero mean and unit variance before being used for neural network training.

\subsection{Model Architecture} 
\subsubsection{Piecewise Polytropic EOS-Based Model\label{sss:piecewisemodel}}
\begin{figure*}[htbp]

        \includegraphics[width=0.6\textwidth]{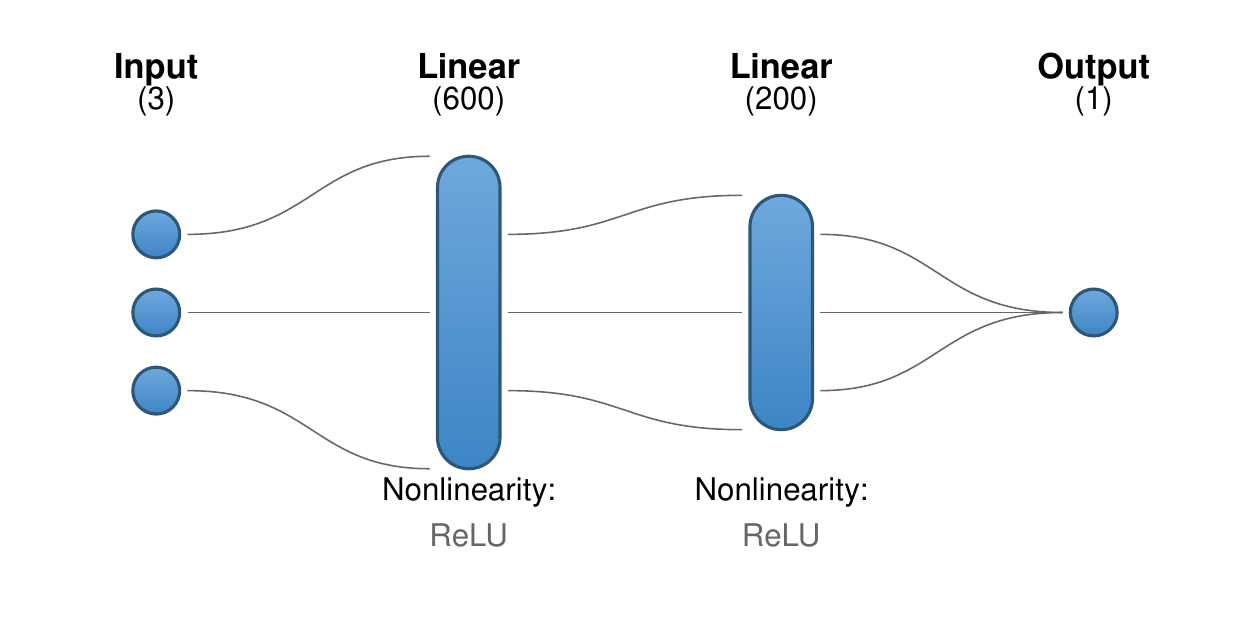}  \\
          \includegraphics[width=0.8\textwidth]{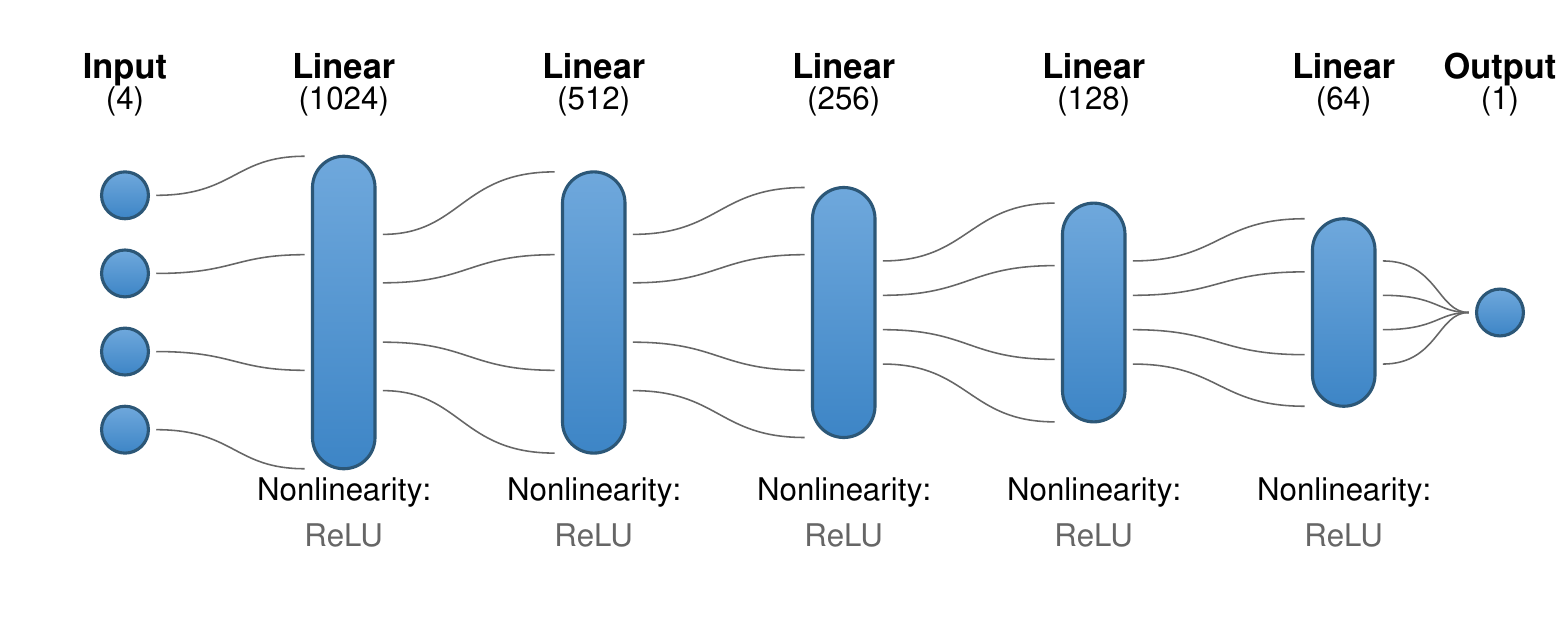}  
     \caption{\justifying Architectures of the neural networks used for conservative-to-primitive variable mapping. Top: The \texttt{NNC2PS} network takes conserved variables $D$, $S_x$, and $\tau$ as input and outputs the pressure $p$. Bottom: The \texttt{NNC2P\_Tabulated} network uses the logarithm of conserved variables $\log D$, $\log S_x$, and $\log \tau$, along with the electron fraction $Y_e$, as input, outputting the logarithm of pressure $\log p$. The \texttt{NNC2PL} network %MDPI: If “Data not shown” is found in the main text, it is required that that this be cited as a reference or deleted directly. Please modify.
     %Author: Fixed. We removed "(not shown)" and replaced it with the highlighted minor modifications below
 shares {an identical} architecture {with} \texttt{NNC2P\_Tabulated}, but with the input/output structure {of} \texttt{NNC2PS}.}
    \label{fig:combined_networks}
\end{figure*}

For the hybrid piecewise polytropic EOS-based model, we tested two feedforward neural networks of varying complexity to represent the conservative-to-primitive variable transformation. Each network takes as input the three conserved variables $(D, S_x, \tau)$ (\mbox{Equation (\ref{eq:cons})}) and outputs the pressure $p$ (Equation (\ref{eq:prim})), assuming the remaining momentum density components are zero for simplicity. This architecture is designed to effectively learn the hidden symmetries in the relationship between the conserved and primitive variables, approximating the intricate C2P transformation without explicit root-finding. After experimenting with multiple multi-layer perceptron (MLP) architectures, as detailed in Appendix~\ref{sec:appendix}, we identified two models that offered an optimal balance between accuracy, speed, and trainability. The smaller model, \texttt{NNC2PS}, features two hidden layers with 600 and 200 neurons, while the larger model, \texttt{NNC2PL}, contains five hidden layers with 1024, 512, 256, 128, and 64 neurons (Figure~\ref{fig:combined_networks}).

ReLU activation functions were applied to the hidden layers to introduce nonlinearity, with the output layer kept linear. We found these models strike an effective balance between complexity and performance, making them well-suited for our task.

\subsubsection{Tabulated EOS-Based Model}
For the tabulated EOS-based model, we use a single feedforward neural network, \texttt{NNC2P\_Tabulated}, to achieve an inherently equivalent task with minor differences. This model takes as input the log-scaled variables $(\log D, \log S_x, \log \tau, \log Y_e)$ and outputs the log-scaled pressure, $\log p$ (Equation (\ref{eq:prim})), assuming $S_y$ and $S_z$ are zero for simplicity as before. Using log-scaled inputs and outputs aligns with the format of the tabulated EOS values, which are also stored in logarithmic form to accommodate the typically large values of these physical quantities. This approach reduces the range of feature magnitudes, facilitating more stable learning dynamics and better alignment with the source data.

We explored several MLP architectures, varying in parameters, layers, and training strategies, to identify an optimal design for our task. Among these, an architecture identical to \texttt{NNC2PL}, featuring five hidden layers with 1024, 512, 256, 128, and 64 neurons, respectively, detailed in Section~\ref{sss:piecewisemodel} above, emerged as a robust choice. This architecture effectively balanced capacity and efficiency, enabling accurate learning of log-scaled pressure from tabulated EOS data (Figure~\ref{fig:combined_networks}).

\subsection{Training Approach}
We use a similar procedure to optimize all neural networks: \texttt{NNC2PS}, \texttt{NNC2PL}, and the tabulated baseline model, \texttt{NNC2P\_Tabulated}, with minor tweaks. Training was performed on a single \texttt{NVIDIA A100} GPU on the Delta cluster. For the hybrid piecewise polytropic EOS-based models (\texttt{NNC2PS} and \texttt{NNC2PL}), we employed a custom, physics-informed loss function that penalizes negative pressure predictions. This loss function is a modified mean-squared error:\vspace{-6pt}
\begin{equation}
\label{eq:loss}
\mathcal{L}(\theta) = \frac{1}{n} \sum_{i=1}^{n} (\hat{y}_i(\theta) - y_i)^2 + q \cdot \sum_{i=1}^{n} \mathrm{ReLU}(-\mathcal{N}^{-1}(\hat{y}_i(\theta))) \,,\vspace{-6pt}
\end{equation}
where $\hat{y}_i(\theta)$ represents the network's estimation for feature $i$, $y_i$ is the corresponding target value, ReLU is the familiar rectified linear unit defined by $\mathrm{ReLU}(x) = \max(0, x)$, and $\mathcal{N}^{-1} (\cdot)$ represents an inverse normalization procedure based on the training data statistics. The penalty factor, $q$, was optimized for each model, with $q=150$ for \texttt{NNC2PS} and $q=350$ for \texttt{NNC2PL}. These values consistently suppressed negative pressure predictions on the test set. For the tabulated EOS model (\texttt{NNC2P\_Tabulated}), the structure of the data precluded negative predictions, so a standard mean-squared error loss function was used.

All models were trained using the Adam optimizer with an initial learning rate of $3 \times 10^{-4}$. A learning rate scheduler reduced the learning rate by a factor of 0.5 if the validation loss failed to improve for five consecutive epochs. \texttt{NNC2PS} and \texttt{NNC2PL} were trained for 85 epochs, while \texttt{NNC2P\_Tabulated} required 250 epochs. These epoch counts were determined empirically by monitoring the validation loss, with training stopped once the loss had clearly converged. The use of a learning rate scheduler, which reduces the learning rate when the validation loss plateaus, also serves as a form of early stopping. For each epoch, the model was set to training mode, and data was loaded in batches of 32 onto the GPU. This batch size was chosen based on experimentation to balance the number of epochs and overall time to convergence. While training with larger batches and multiple GPUs (using  PyTorch's \texttt{DataParallel} module or other approaches) is possible, we found no significant advantage regarding the total time to convergence and ultimately opted for this simpler, more portable approach. For each batch, optimizer gradients were reset before generating predictions, and the loss was computed using respective loss functions. Backpropagation was then performed to update the model parameters.

After completing the training phase for each epoch, the model's performance is evaluated on the validation dataset, accumulating the validation loss similarly to the training loss. Both losses are normalized by the size of the respective datasets and stored for further analysis, specifically for clues of potential overtraining.

\subsection{Inference Speed Tests}
In our inference speed tests, we evaluated two main approaches for efficient deployment: a TorchScript model and NVIDIA’s TensorRT optimized engines. These tests were conducted to measure and compare inference speed under typical deployment conditions, aiming to take advantage of the \texttt{A100} GPU on Delta.

\subsubsection{TorchScript Deployment}
To prepare models for inference with TorchScript, we first saved a scripted version of the model, which is compatible with PyTorch’s JIT compiler, optimizing runtime execution without modifying the model’s core structure. TorchScript’s scripting provides some degree of optimization, enabling faster model execution than standard PyTorch models but without the hardware-level optimizations that TensorRT offers.

\subsubsection{TensorRT Deployment}
For TensorRT, we explored both \texttt{FP32} (unquantized) and \texttt{FP16}-quantized engines, ultimately deciding not to pursue \texttt{INT8} quantization due to accuracy degradation observed in initial tests. After extensive testing, we opted for dynamic engine building with a batch size determined by the total size of the expected dataset, as this approach provided the best balance between performance and flexibility for our hardware and model structure. It must be noted that constructing an optimal engine in TensorRT is a nuanced process, influenced by multiple factors including model architecture, hardware specifications, intended batch sizes during inference, and input data. Therefore, achieving the best results often involves iterative tuning and profiling to adapt the engine to the specific deployment environment and workload requirements. Below, we summarize the overall engine-building process we followed in detail:

\begin{itemize}
    \item \textbf{{Model Export to ONNX:} %MDPI: Please confirm if the bold formatting is necessary; if not, please remove it. The following highlights are the same.
    %Author: We confirm the bold formatting in this particular instance is necessary.
} First, we exported the PyTorch model to the ONNX format. This conversion enables interoperability with TensorRT, which uses ONNX as its primary model input format. 

    \item \textbf{{TensorRT Engine Building:}} Using TensorRT’s {Python} %MDPI: Please state the version number of the software.
    %Author: This information is directly tied to the PyTorch and TensorRT versions, which have already been clarified above; therefore, no update is necessary.
 API, we constructed both \texttt{FP32} and \texttt{FP16} engines. A logger was initialized for verbose logging to capture potential issues during engine building. With the TensorRT \texttt{Builder}, we created a network definition with explicit batch handling, which is essential for dynamic batching configurations.

    \item \textbf{{Parsing and Validating the ONNX Model:}} We loaded the ONNX model into TensorRT, where the \texttt{OnnxParser} validated and parsed the model. Parsing errors, if any, were logged for troubleshooting, ensuring a valid model structure before optimization.

    \item \textbf{{Configuration and Optimization Profiles:}} The \texttt{BuilderConfig} was set with a 40 GB workspace memory limit, providing more than enough headroom for dynamic batch sizes while maintaining stable performance. We set up a dynamic optimization profile specifying minimum, optimal, and maximum batch sizes within a 10 percent margin of our typical usage, granting flexibility to handle both smaller and larger input \mbox{volumes efficiently. }

    \item \textbf{{Engine Serialization:}} Finally, we serialized and saved the engine, creating a portable and optimized binary that can be loaded for deployment. This step encapsulates the model’s architecture, weights, and optimizations, ensuring it is ready for \mbox{fast inference.}
\end{itemize}

To ensure we measure the maximum possible performance for each point in our benchmark, we build a specialized, yet flexible, TensorRT engine for each combination of model and dataset size. The dynamic optimization profile for each of these engines is configured with a tight margin around its target dataset size ($N$), as detailed in Table~\ref{tab:tensorrt_profiles}.

\begin{table*}[t]
\caption{Dynamic optimization profiles used for building specialized TensorRT engines for each benchmarked dataset size (N). Each profile is configured with a tight margin around its target optimal size.}
\label{tab:tensorrt_profiles}
\centering
\begin{tabular}{cccc}
\toprule
\textbf{Target Dataset Size (N)} & \textbf{Min Batch Size (0.95 N)} & \textbf{Optimal Batch Size (N)} & \textbf{Max Batch Size (1.05 N)} \\
\midrule
25,000 & 23,750 & 25,000 & 26,250 \\
50,000 & 47,500 & 50,000 & 52,500 \\
100,000 & 95,000 & 100,000 & 105,000 \\
500,000 & 475,000 & 500,000 & 525,000 \\
1,000,000 & 950,000 & 1,000,000 & 1,050,000 \\
\bottomrule
\end{tabular}
\end{table*}

Overall, the process of optimizing and saving models using both TorchScript and TensorRT gave us insight into balancing flexibility, accuracy, and performance. For larger batch sizes and greater computational demands, TensorRT’s dynamic engine approach in \texttt{FP16} is often more effective, even for models as simple as ours, while TorchScript remains a reliable fallback and simpler alternative.

For the actual inference speed test procedure, we implemented two distinct workflows on a single GPU for both approaches. The TorchScript-based approach allowed for a straightforward configuration, primarily requiring the definition of batch sizes and the pre-loading of data onto the GPU. It then used \texttt{libtorch} for efficient GPU deployment and batch execution.

In contrast, the TensorRT-based approach demanded several additional configurations. The model, after being converted into an optimized engine, was loaded using TensorRT's \texttt{C++} API. This included the manual pre-loading of input data into GPU memory before execution and was followed by manual setup of input and output buffers for TensorRT's \texttt{executeV2} function and careful management of CUDA resources. While this setup was more involved, it leveraged hardware-specific optimizations to deliver substantial gains in inference speed.

\section{\label{sec:res}Results}
\subsection{Accuracy}

We evaluate the model accuracy using two standard metrics for regression problems: the $L_1$ error (mean absolute error) and the $L_{\infty}$ error (maximum absolute error), both calculated over the entire test dataset. Table \ref{tab:accuracy_results} summarizes the accuracy results based on $L_1$ and $L_{\infty}$ error metrics for each model variant—\texttt{NNC2PS}, \texttt{NNC2PL}, and \texttt{NNC2P\_Tabulated}—including both the unquantized and quantized TensorRT engines built from them. 

The \texttt{NNC2PS} model trained in PyTorch achieves very high accuracy with an $L_1$ error of $4.54 \times 10^{-7}$ and an $L_{\infty}$ error of $3.44 \times 10^{-6}$. When the model is converted to a TensorRT engine, the accuracy remains nearly identical, with an $L_1$ error of $4.54 \times 10^{-7}$ and an $L_{\infty}$ error of $3.43 \times 10^{-6}$, indicating minimal loss in precision due to TensorRT optimization. However, when \texttt{FP16} quantization is applied, the error rates increase to an $L_1$ error of $6.39 \times 10^{-7}$ and an $L_{\infty}$ error of $8.98 \times 10^{-6}$, revealing an obvious side effect of reduced precision. This highlights the classic trade-off between computational performance and numerical precision, a critical consideration for selecting the appropriate model for a given scientific application where the tolerance for numerical error may vary.

\begin{table*}[htbp]
\centering
\caption{Accuracy results for all models.}
\setlength\tabcolsep{0.805cm}
\begin{tabular}{lccc}
 \toprule
\textbf{Model} & \boldmath$L_1$ \textbf{Error} & \boldmath$L_{\infty}$ \textbf{Error} \\
\hline
\texttt{NNC2PS} (PyTorch) & $4.54 \times 10^{-7}$ & $3.44 \times 10^{-6}$ \\
\texttt{NNC2PS} (TensorRT) & $4.54 \times 10^{-7}$ & $3.43 \times 10^{-6}$ \\
\texttt{NNC2PS} (TensorRT--\texttt{FP16}) & $6.39 \times 10^{-7}$ & $8.98 \times 10^{-6}$ \\
\hline
\texttt{NNC2PL} (PyTorch) & $2.75 \times 10^{-7}$ & $2.61 \times 10^{-6}$ \\
\texttt{NNC2PL} (TensorRT) & $2.88 \times 10^{-7}$ & $2.69 \times 10^{-6}$ \\
\texttt{NNC2PL} (TensorRT--\texttt{FP16}) & $5.32 \times 10^{-7}$ & $9.84 \times 10^{-6}$ \\
\hline
\texttt{NNC2P\_Tabulated} (PyTorch) & $8.02 \times 10^{-3}$ & $3.54 \times 10^{-1}$ \\
\texttt{NNC2P\_Tabulated} (TensorRT) & $8.16 \times 10^{-3}$ & $3.45 \times 10^{-1}$ \\
\texttt{NNC2P\_Tabulated} (TensorRT--\texttt{FP16}) & $1.38 \times 10^{-2}$ & $7.44 \times 10^{-1}$ \\
 \bottomrule
\end{tabular}
\label{tab:accuracy_results}
\end{table*}

The larger \texttt{NNC2PL} model, rather expectedly, achieves lower $L_1$ and $L_{\infty}$ errors than \texttt{NNC2PS}, with an $L_1$ error of $2.75 \times 10^{-7}$ and an $L_{\infty}$ error of $2.61 \times 10^{-6}$. The corresponding TensorRT engine preserves this high level of accuracy, showing only a slight and negligible increase to an $L_1$ error of $2.88 \times 10^{-7}$ and $L_{\infty}$ error of $2.69 \times 10^{-6}$, respectively. The \texttt{FP16} quantized version, however, sees a notable rise in error metrics, with an $L_1$ error of \mbox{$5.32 \times 10^{-7}$} and an $L_{\infty}$ error of $9.84 \times 10^{-6}$.

The \texttt{NNC2P\_Tabulated} model exhibits an $L_1$ error of $8.02 \times 10^{-3}$ and an $L_{\infty}$ error of $3.54 \times 10^{-1}$. It is important to clarify that this larger error does not indicate a failure of the ML model but is a direct consequence of the model learning from a completely different dataset constructed from the LS220 EOS table to estimate the logarithmic pressure values. The TensorRT engine version also shows only a slight increase in $L_1$ error to $8.16 \times 10^{-3}$. With \texttt{FP16} quantization, the $L_1$ error rises again, more noticeably, to $1.38 \times 10^{-2}$.

\begin{figure*}[htbp]

    \includegraphics[width=1.0\textwidth]{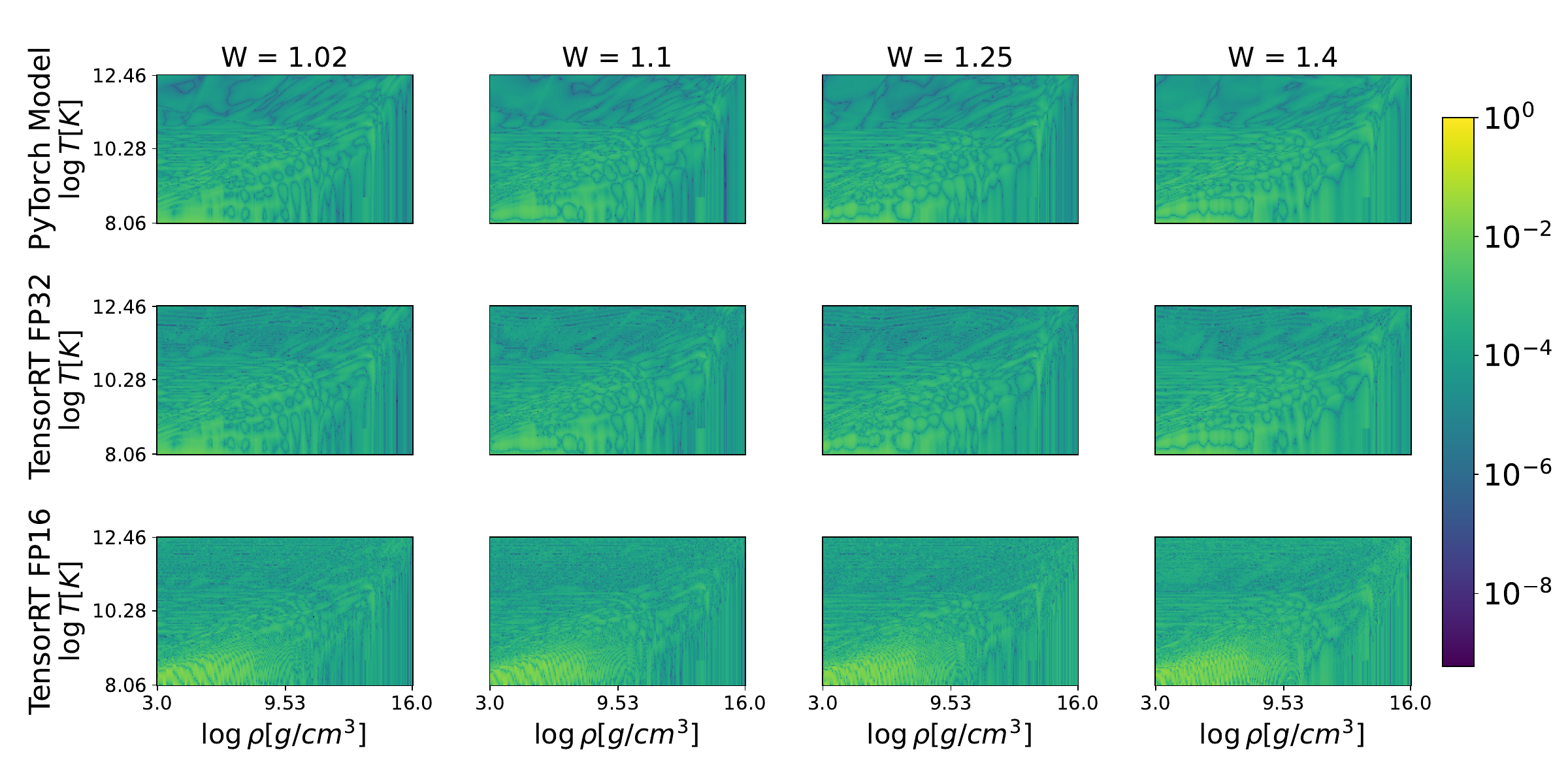}
    \caption{\justifying Relative error of the \texttt{NNC2P\_Tabulated} model for various Lorentz factors ($W$) with $Y_e \approx 0.1$. The plots highlight the accuracy trends across different regions of the \texttt{LS220} EOS table, showing larger relative errors in low-density and low-temperature regions, reflecting the inherent complexities of the EOS in this region. This behavior is consistent across the tested $W$ values of 1.02, 1.1, 1.25, and 1.4 and is more pronounced for the \texttt{FP16} precision TensorRT engine.}
    \label{fig:tab_comparison}

\end{figure*}

Additionally, we examined the relative accuracy of the \texttt{NNC2P\_Tabulated} model for parameters $W = 1.02$, $1.1$, $1.25$, and $1.4$ with $Y_e \approx 0.1$ (See Figure~\ref{fig:tab_comparison}). The relative error, defined as the absolute error divided by the true value for each point in a specific parameter set, was not uniform across the parameter space. Larger relative errors were observed in the lowest density and temperature regions of the EOS table, while slightly smaller errors occurred in the high-temperature regions. This accuracy trend was consistent across all tested Lorentz factor ($W$) values and even more emphasized for the \texttt{FP16} precision TensorRT engine. The \texttt{LS220} EOS, as provided by ~\cite{SRO2017}, transitions from detailed treatment at high densities to simplified approximations at lower densities, which may contribute to these disparities. Low-density regions are inherently challenging due to the dominance of thermal effects, non-uniform phase transitions, and the treatment of nuclear matter surfaces, which can exacerbate modeling errors~\cite{SRO2017, bernuzzi2020}. These characteristics likely explain the reduced accuracy in these regions, where variations in the nuclear matter’s phase state are more pronounced.

The overall results show that TensorRT’s optimizations maintain accuracy across models when using full precision. \texttt{FP16} quantization, while accelerating inference (as will be discussed further below), introduces higher error rates, particularly in certain models. The potential trade-off between the inference speed and precision can be especially important in relativistic hydrodynamics simulations, where the accuracy of small-scale structures and wave propagation can critically impact the fidelity of predictions. For such simulations, even slight deviations due to quantization can influence results, making full-precision TensorRT inference particularly valuable when accuracy is paramount. Conversely, \texttt{FP16} quantization may be suitable for faster, lower-fidelity simulations where minor accuracy trade-offs are acceptable.

\subsection{Inference Speed Analysis}

The inference performance of various methods was evaluated using a single NVIDIA A100 GPU for neural network models and a single-threaded CPU implementation of the traditional numerical method from the RePrimAnd library. The CPUs used in this study were dual AMD 64-core 2.45 GHz Milan processors on the Delta cluster, which can support up to 128 threads. Each configuration was tested across five dataset sizes, ranging from 25,000 to 1,000,000 data points, with ten inference runs conducted per configuration to ensure result stability and consistency. For the RePrimAnd numerical solver, we set the target accuracy for the relative error in the root-finding algorithm to $10^{-8}$. This is a standard, high-precision value used in production codes. We chose to compare our ML models against this robust baseline rather than tuning the numerical solver's accuracy to match that of the NNs, ensuring a conservative performance comparison.

\begin{figure*}[htbp]
    \centering
   \includegraphics[width=1.0\textwidth]{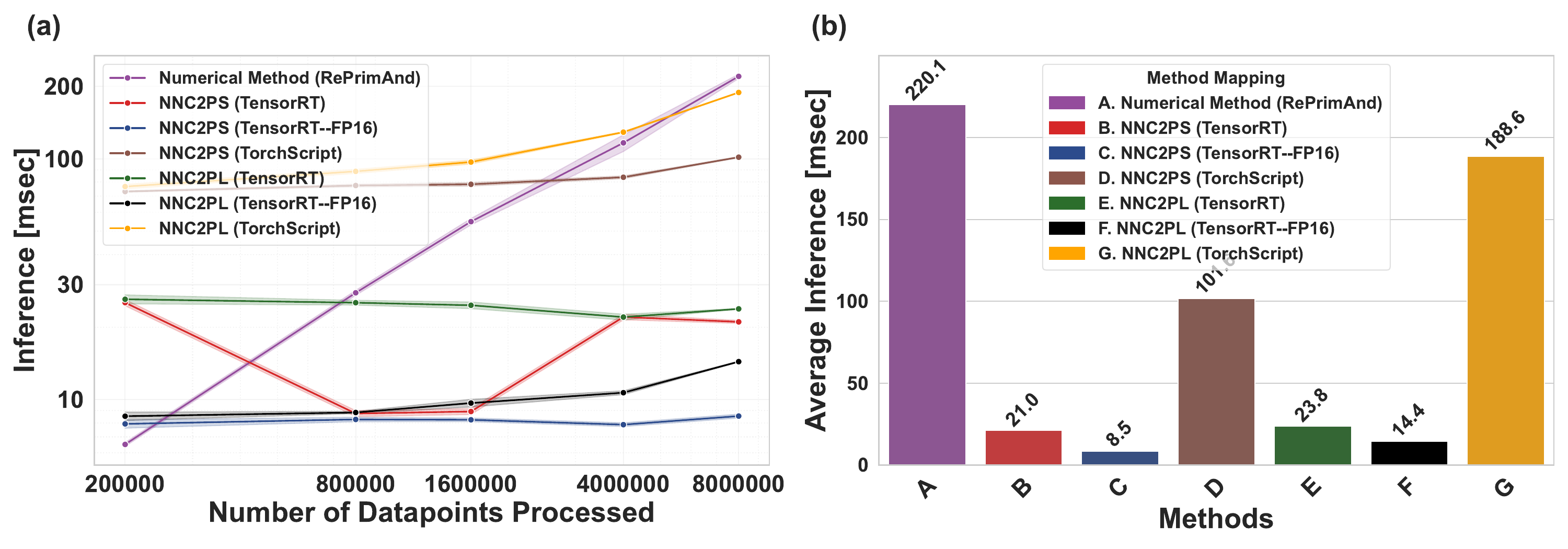}
    \caption{\justifying Ideal scaling comparison of various C2P inversion methods under the assumption of perfect parallelization. (\textbf{a}) Projected inference time as a function of dataset size for a traditional numerical solver (RePrimAnd utilizing 128 CPU threads on a single node of the Delta cluster) and two neural network models (\texttt{NNC2PS} and \texttt{NNC2PL}) using TensorRT (\texttt{FP32} and \texttt{FP16} precision) and TorchScript across 8 NVIDIA A100 GPUs. (\textbf{b}) Projected inference speed comparison for a dataset of 8 million points, highlighting the significant scalability and efficiency gains achieved by TensorRT engines, particularly with FP16 optimization. The mixed-precision TensorRT engine for \texttt{NNC2PS} achieves approximately a 25-fold reduction in processing time compared to the numerical method, showcasing the potential for TensorRT-based methods to convincingly outperform traditional numerical solvers at scale. The width of the lines in panel (\textbf{a}) represents the standard deviation over ten independent runs, with the wider band for the numerical method indicating higher runtime variability.}
    \label{fig:performance}

\end{figure*}

The numerical method exhibited linear scaling of inference time with respect to the dataset size. In contrast, both TensorRT and TorchScript models generally maintained relatively stable inference times across the dataset sizes. Notably, the full-precision TensorRT engine for the smaller network, \texttt{NNC2PS}, showed a faster-than-expected processing time at certain intermediate dataset sizes, as observed in Figure~\ref{fig:performance}a. This behavior may be attributed to favorable thread block utilization and the kernel selection mechanism of TensorRT for this particular network size. A more detailed profiling study is needed to fully elucidate the underlying cause. The accuracy characteristics of these models remained consistent, as indicated in Table~\ref{tab:accuracy_results}.

The numerical method required significantly more time than the neural network-based approaches. On average, the numerical method took 103.8 ms to process \mbox{25,000 data} points, with runtime scaling almost linearly to 3490 ms for 1,000,000 data points. In contrast, the neural network models demonstrated substantially faster inference times. Specifically, the mixed-precision TensorRT engine built from \texttt{NNC2PS} required 7.92 ms for 25,000 data points and 8.54 ms for 1,000,000 data points. Its full-precision counterpart exhibited similar performance, with runtimes of 25.17 ms for 25,000 data points and \mbox{21.06 ms} for \mbox{1,000,000 data} points. The TorchScript variant showed slower performance but still maintained sub-linear scaling, with runtimes averaging 72.79 ms for 25,000 points and 101.74 ms for 1,000,000 points.\vspace{-3pt}

A similar trend was observed for the \texttt{NNC2PL} models, with TensorRT engines consistently outperforming their TorchScript counterparts. The mixed-precision TensorRT engine for \texttt{NNC2PL} processed 25,000 data points in 8.32 ms and 1,000,000 points in 14.35 ms. In comparison, the full-precision TensorRT engine required 25.85 ms for 25,000 points and 23.87 ms for 1,000,000 points. The TorchScript model averaged 73.18 ms for 25,000 points and 102.04 ms for 1,000,000 points.

Figure~\ref{fig:performance} presents a theoretical performance benchmark based on ideal scaling under the assumption of perfect parallelization. This scenario assumes optimal workload distribution, minimal communication overhead, and negligible synchronization delays, representing the upper bound of scalability. For the numerical method, the figure reflects the full computational capacity of a single CPU node on the Delta cluster, utilizing \mbox{128 threads}. For the neural networks, it represents the use of 8 A100 GPUs within a single GPU node. Under these ideal conditions, the processing time of the numerical method per data point is projected to decrease by a factor of 128, allowing for the processing of 8 million points in approximately 218 ms (Figure~\ref{fig:performance}b). Similarly, all neural network methods are expected to achieve linear inference scaling with similar per-GPU efficiency. Under this scenario, TensorRT-based methods—particularly the mixed-precision engine for \texttt{NNC2PS}—show a 25-fold reduction in processing time for 8 million points compared to the numerical method running at full capacity on the CPU node. Furthermore, the scaling trend strongly favors TensorRT for even larger datasets.

The results presented above underscore the substantial performance gains achievable through the use of TensorRT-optimized neural networks, particularly in the context of conservative-to-primitive inversion in relativistic hydrodynamics simulations. By leveraging the parallel processing power of modern GPUs, these methods offer significant speedups compared to traditional CPU-based numerical approaches, even in large-scale simulations involving millions of data points. As demonstrated, TensorRT optimizations enable more efficient and scalable solutions, with the potential to dramatically reduce the computational cost of C2P operations. This work highlights the clear advantage of integrating ML-driven methods with GPU acceleration to address the computational challenges of high-throughput simulations. Moving forward, the next step is to incorporate these optimized approaches into full-scale hydrodynamics simulations, where their impact on both performance and scalability can be fully realized.

It is important to contextualize the comparison between the fully utilized CPU component (128 threads) and the fully utilized GPU component (8 GPUs) of a single compute node. This 'node-to-node' benchmark is designed to answer the practical question of how to best utilize the co-located and often cost-equivalent hardware resources of a modern heterogeneous compute node. While a formal cost-normalized analysis is complex, this approach compares the optimal-use scenario for each hardware type available to a researcher on a typical allocation. The resulting 25-fold speedup is therefore a combination of the algorithmic shift (from iterative root-finding to direct-mapping) and the architectural advantage of GPUs for the massively parallel workload presented by the neural network.

\section{\label{sec:end}Conclusions}

This work introduces a novel ML-driven method for accelerating C2P inversions in relativistic hydrodynamics simulations, with a focus on hybrid piecewise polytropic and tabulated equations of state. By employing feedforward neural networks optimized with TensorRT, we achieve substantial performance improvements over traditional CPU solvers, offering a compelling alternative to computationally expensive iterative methods while maintaining high accuracy. Our results demonstrate that the TensorRT-optimized neural networks can process large datasets significantly faster, achieving up to 25 times the inference speed of traditional methods. The success of this approach is rooted in the neural network's ability to efficiently learn and represent the inherent symmetries and complex functional relationships within the EOS, effectively creating a direct mapping that bypasses iterative numerical solvers.

Future work will explore several key directions to refine and expand this approach. First, adapting the models to handle a broader range of equations of state will improve the versatility of this method across different simulation contexts. Second, exploring alternative network architectures, such as those incorporating physics-informed layers or adaptive activation functions to better handle physical discontinuities like phase transitions, could further enhance both accuracy and inference speed. Third, the models must be extended to handle full three-dimensional velocities to be fully integrated into production-level GRMHD codes. Additionally, continued optimization of TensorRT, including advanced parallelization strategies and scaling across multiple GPUs, and careful exploration of lower-precision formats like INT8, potentially with quantization-aware training, promises even greater reductions in computational time, enabling simulations of larger and more complex astrophysical systems. These improvements will be critical for advancing high-resolution simulations in numerical relativistic hydrodynamics. 

We believe that ML-driven methods, particularly those incorporating TensorRT optimization, will play an essential role in advancing the field of general relativistic hydrodynamics and numerical relativity more broadly. To facilitate further validation and extension of these findings, we have made the software developed for this study publicly available at: %the following \hl{GitHub} %MDPI: Please state the version number of the software.
%Author: GitHub is a website, so not applicable. We have removed the reference to it and added a direct link to the repository (highlighted)
 {\texttt{https://github.com/semihkacmaz/C2PNets}} (accessed on 27 August 2025). %MDPI: Please add the access date (format: Date Month Year), e.g., accessed on 1 January 2020.
 %Author: Added. Thanks!

%=================================================================
% APPENDIX
%=================================================================
\newpage
\appendix
\section{Model Architecture Exploration and Training History}
\label{sec:appendix}

In this study, we explored a wide range of multi-layer perceptron (MLP) architectures to identify models that offer an optimal balance between predictive accuracy and inference speed. The models presented in the main text---\texttt{NNC2PS}, \texttt{NNC2PL}, and \texttt{NNC2P\_Tabulated}---were the result of this systematic exploration.

Our findings, summarized in Table~\ref{tab:arch_exploration}, demonstrate a clear ``sweet spot'' for model complexity. Architectures smaller than our chosen models (e.g., \texttt{NNC2PS-Tiny}) offered lower parameter counts but with a notable drop in accuracy. Conversely, models that were significantly wider or deeper than our selections provided only marginal accuracy gains for a substantial increase in parameter count and computational cost (e.g., \texttt{NNC2PL-Wide}, \texttt{NNC2PL-Deep-7}). This trend of diminishing returns is evident across both EOS types.

Notably, excessively deep architectures (e.g., the 10- and 13-layer models) consistently exhibited training instability or yielded worse performance, reinforcing our choice of moderately sized networks as the most effective and efficient solution for this \mbox{regression task}.

\begin{figure*}[htbp]
\centering
% --- Panel (a) ---
\begin{minipage}[t]{0.48\linewidth}
    \centering
    \includegraphics[width=\linewidth]{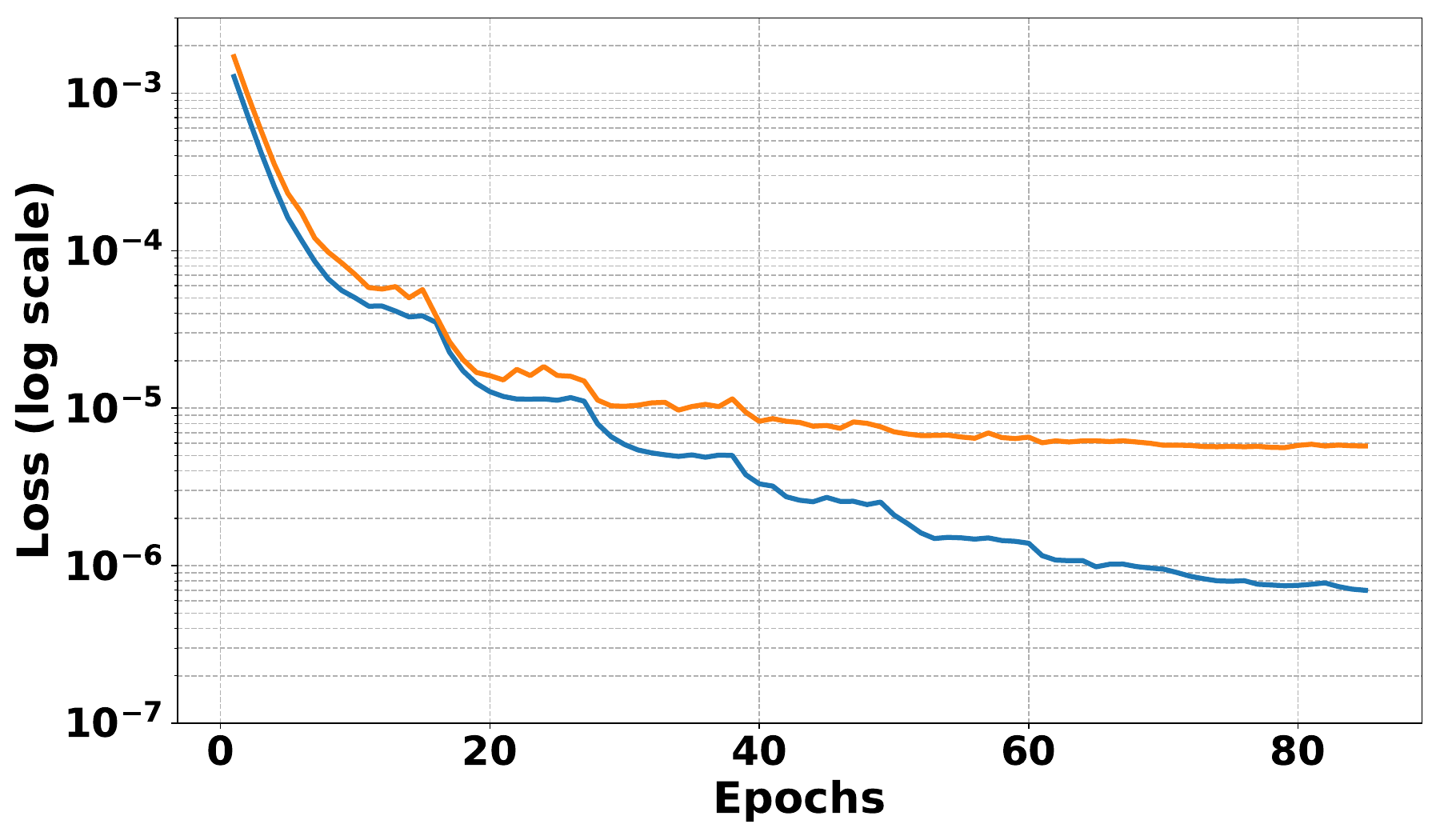}
    \vspace{1mm}
    (\textbf{a}) \texttt{NNC2PS}
\end{minipage}%
\hfill
% --- Panel (b) ---
\begin{minipage}[t]{0.48\linewidth}
    \centering
    \includegraphics[width=\linewidth]{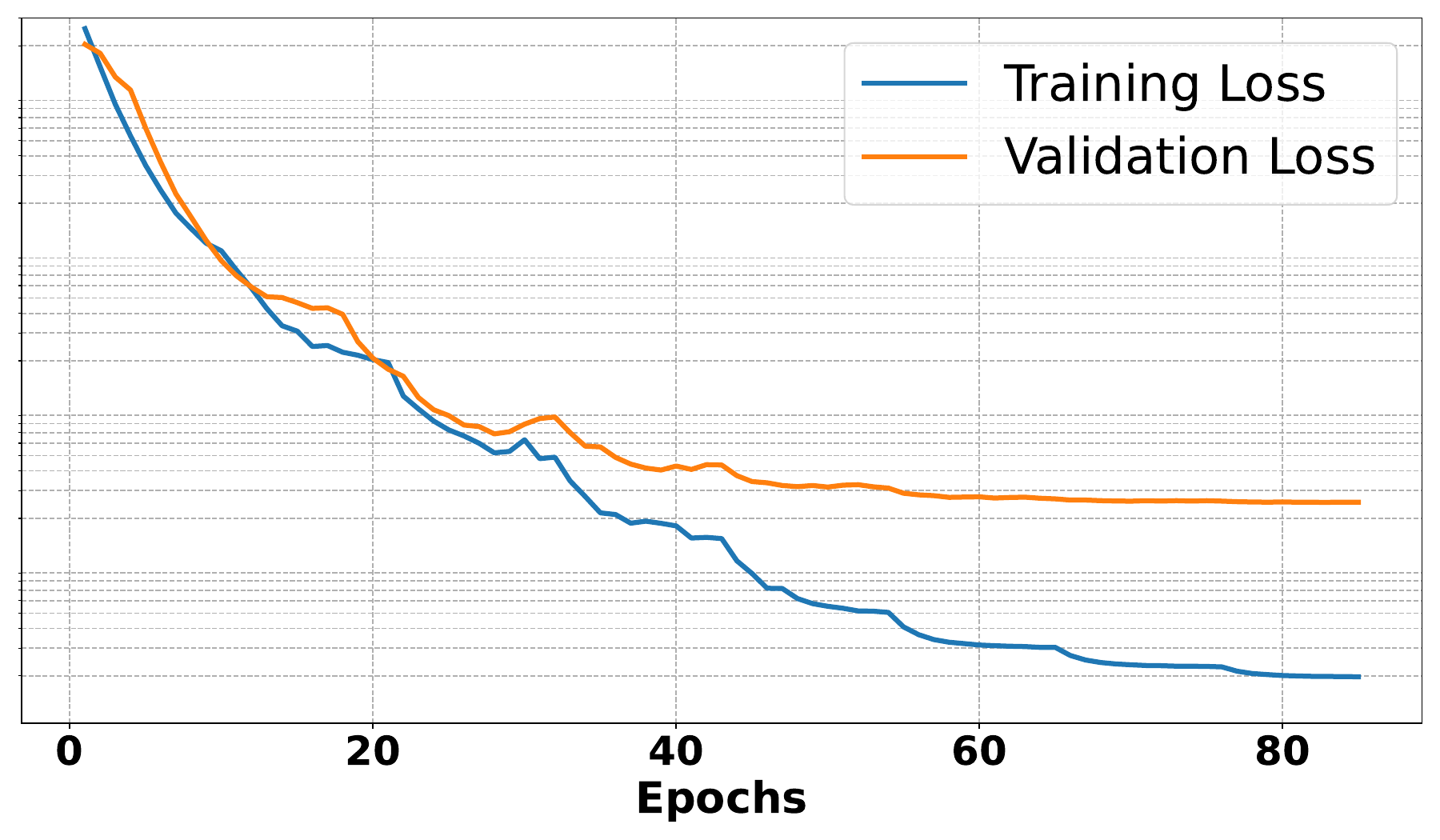}
    (\textbf{b}) \texttt{NNC2PL}
\end{minipage}
\caption{Training and validation loss curves for the piecewise polytropic EOS models. The smooth convergence demonstrates a stable training process for (\textbf{a}) \texttt{NNC2PS} and (\textbf{b}) \texttt{NNC2PL}.}
\label{fig:loss_curves_polytropic}
\end{figure*}

\begin{figure*}[htbp]
\centering
\includegraphics[width=0.6\linewidth]{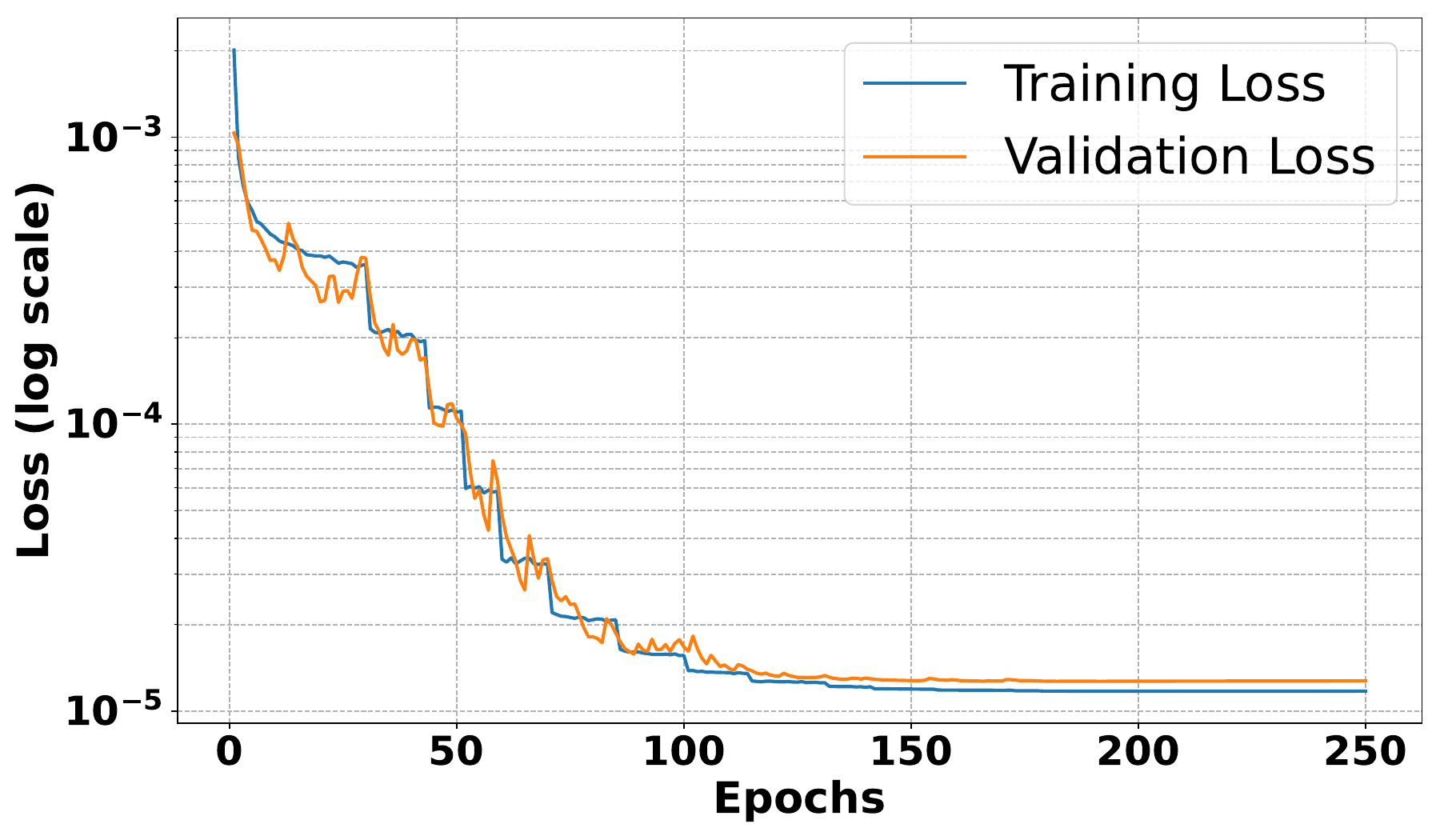}
\caption{Training and validation loss curves for the tabulated EOS model, \texttt{NNC2P\_Tabulated}. The smooth convergence demonstrates a stable training process without significant overfitting.}
\label{fig:loss_curve_tabulated}
\end{figure*}

To demonstrate the stability of our training procedure, Figures~\ref{fig:loss_curves_polytropic} and~\ref{fig:loss_curve_tabulated} show the training and validation loss curves for the final three models used in this work. The curves illustrate smooth convergence to a low loss value with no signs of significant overfitting, validating our training methodology.

\begin{table*}
\caption{Explored architectures and validation accuracy ($L_1$ error) for both EOS models. The selected models are shown in bold. The validation error measures the model's performance on unseen data.}
\label{tab:arch_exploration}
\centering
\begin{tabular}{llcc}
\toprule
\textbf{Model Name} & \textbf{Hidden Layers (Neurons per Layer)} & \textbf{Total Parameters} & \textbf{Validation $L_1$ Error} \\
\midrule
\multicolumn{4}{c}{\textbf{Piecewise Polytropic EOS}} \\
\midrule
\texttt{NNC2PS-Tiny} & [300, 100] & \textasciitilde31 k & $5.8 \times 10^{-7}$ \\
\texttt{NNC2PS-Shallow} & [800] & \textasciitilde3 k & $6.5 \times 10^{-7}$ \\
\textbf{\texttt{NNC2PS}} & \textbf{[600, 200]} & \textbf{\textasciitilde123 k} & $\mathbf{4.5 \times 10^{-7}}$ \\
\texttt{NNC2PS-Wide} & [800, 400] & \textasciitilde324 k & $4.1 \times 10^{-7}$ \\
\midrule
\texttt{NNC2PL-Small} & [512, 256, 128, 64] & \textasciitilde180 k & $3.9 \times 10^{-7}$ \\
\texttt{NNC2PL-Medium} & [1024, 512, 256, 128] & \textasciitilde690 k & $3.2 \times 10^{-7}$ \\
\textbf{\texttt{NNC2PL}} & \textbf{[1024, 512, 256, 128, 64]} & \textbf{\textasciitilde707 k} & $\mathbf{2.8 \times 10^{-7}}$ \\
\texttt{NNC2PL-Wide} & [2048, 1024, 512, 256, 128] & \textasciitilde2.8 M & $2.5 \times 10^{-7}$ \\
\texttt{NNC2PL-Deep-7} & [1024, 1024, 512, 512, 256, 128, 64] & \textasciitilde2.4 M & $2.9 \times 10^{-7}$ \\
\texttt{NNC2PL-Deep-10} & [1024, 1024, 512, 512, 256, 256, 128, 128, 64, 64] & \textasciitilde3.5 M & $3.1 \times 10^{-7}$ \\
\texttt{NNC2PL-SuperDeep} & 13 Layers & \textasciitilde5 M & Failed to Converge \\
\midrule
\multicolumn{4}{c}{\textbf{Tabulated EOS (LS220)}} \\
\midrule
\texttt{NNC2P\_Tab-Tiny} & [512, 256, 128] & \textasciitilde165 k & $9.5 \times 10^{-3}$ \\
\texttt{NNC2P\_Tab-Small} & [1024, 512, 256, 128] & \textasciitilde690 k & $8.8 \times 10^{-3}$ \\
\textbf{\texttt{NNC2P\_Tabulated}} & \textbf{[1024, 512, 256, 128, 64]} & \textbf{\textasciitilde707 k} & $\mathbf{8.0 \times 10^{-3}}$ \\
\texttt{NNC2P\_Tab-Wide} & [2048, 1024, 512, 256, 128] & \textasciitilde2.8 M & $7.7 \times 10^{-3}$ \\
\texttt{NNC2P\_Tab-Deep-7} & [1024, 1024, 512, 512, 256, 128, 64] & \textasciitilde2.4 M & $8.2 \times 10^{-3}$ \\
\texttt{NNC2P\_Tab-Deep-10} & [1024, 1024, 512, 512, 256, 256, 128, 128, 64, 64] & \textasciitilde3.5 M & $8.5 \times 10^{-3}$ \\
\texttt{NNC2P\_Tab-SuperDeep} & 13 Layers & \textasciitilde5 M & Failed to Converge \\
\bottomrule
\end{tabular}
\end{table*}

\clearpage

% \bibliographystyle{IEEEtran}
% \bibliography{manuscript}% Produces the bibliography via BibTeX.

\begin{thebibliography}{999}

\bibitem[Radice et~al.(2020)Radice, Bernuzzi, and Perego]{Radice_2020}
Radice, D.; Bernuzzi, S.; Perego, A.
\newblock The Dynamics of Binary Neutron Star Mergers and GW170817.
\newblock {\em Annu. Rev. Nucl. Part. Sci.} {\bf 2020}, {\em
  70},~95–119.
\newblock {\url{https://doi.org/10.1146/annurev-nucl-013120-114541}}.

\bibitem[Ciolfi et~al.(2017)Ciolfi, Kastaun, Giacomazzo, Endrizzi, Siegel, and
  Perna]{Ciolfi:2017uak}
Ciolfi, R.; Kastaun, W.; Giacomazzo, B.; Endrizzi, A.; Siegel, D.M.; Perna, R.
\newblock {General relativistic magnetohydrodynamic simulations of binary
  neutron star mergers forming a long-lived neutron star}.
\newblock {\em Phys. Rev. D} {\bf 2017}, {\em 95},~063016.
\newblock {\url{https://doi.org/10.1103/PhysRevD.95.063016}}.

\bibitem[Kiuchi(2024)]{Kiuchi:2024lpx}
Kiuchi, K.
\newblock {General relativistic magnetohydrodynamics simulations for binary
  neutron star mergers.} \emph{arXiv} {\bf 2024}, arXiv:2405.10081.


\bibitem[Siegel and Metzger(2017)]{PhysRevLett.119.231102}
Siegel, D.M.; Metzger, B.D.
\newblock Three-Dimensional General-Relativistic Magnetohydrodynamic
  Simulations of Remnant Accretion Disks from Neutron Star Mergers: Outflows
  and $r$-Process Nucleosynthesis.
\newblock {\em Phys. Rev. Lett.} {\bf 2017}, {\em 119},~231102.
\newblock {\url{https://doi.org/10.1103/PhysRevLett.119.231102}}.

\bibitem[Sun et~al.(2022)Sun, Ruiz, Shapiro, and Tsokaros]{Sun:2022vri}
Sun, L.; Ruiz, M.; Shapiro, S.L.; Tsokaros, A.
\newblock {Jet launching from binary neutron star mergers: Incorporating
  neutrino transport and magnetic fields}.
\newblock {\em Phys. Rev. D} {\bf 2022}, {\em 105},~104028.
 \newblock {\url{https://doi.org/10.1103/PhysRevD.105.104028}}.

\bibitem[Tsokaros et~al.(2022)Tsokaros, Ruiz, Shapiro, and
  Ury\={u}]{Tsokaros:2021pkh}
{Tsokaros, A.; Ruiz, M.; Shapiro, S.L.; Ury\={u}, K.
\newblock Magnetohydrodynamic Simulations of Self-Consistent Rotating Neutron
  Stars with Mixed Poloidal and Toroidal Magnetic Fields.
\newblock {\em Phys. Rev. Lett.} {\bf 2022}, {\em 128},~061101.
\newblock {\url{https://doi.org/10.1103/PhysRevLett.128.061101}}}.

\bibitem[Fern{\'a}ndez et~al.(2019)Fern{\'a}ndez, Tchekhovskoy, Quataert,
  Foucart, and Kasen]{fernandez2019long}
Fern{\'a}ndez, R.; Tchekhovskoy, A.; Quataert, E.; Foucart, F.; Kasen, D.
\newblock Long-term GRMHD simulations of neutron star merger accretion discs:
  implications for electromagnetic counterparts.
\newblock {\em Mon. Not. R. Astron. Soc.} {\bf 2019},
  {\em 482},~3373--3393.

\bibitem[Foucart et~al.(2016)Foucart, Haas, Duez, O’Connor, Ott, Roberts,
  Kidder, Lippuner, Pfeiffer, and Scheel]{foucart2016low}
Foucart, F.; Haas, R.; Duez, M.D.; O’Connor, E.; Ott, C.D.; Roberts, L.;
  Kidder, L.E.; Lippuner, J.; Pfeiffer, H.P.; Scheel, M.A.
\newblock Low mass binary neutron star mergers: Gravitational waves and
  neutrino emission.
\newblock {\em Phys. Rev. D} {\bf 2016}, {\em 93},~044019.

\bibitem[Camilletti et~al.(2022)Camilletti, Chiesa, Ricigliano, Perego,
  Lippold, Padamata, Bernuzzi, Radice, Logoteta, and
  Guercilena]{Camilletti:2022jms}
Camilletti, A.; Chiesa, L.; Ricigliano, G.; Perego, A.; Lippold, L.C.;
  Padamata, S.; Bernuzzi, S.; Radice, D.; Logoteta, D.; Guercilena, F.M.
\newblock {Numerical relativity simulations of the neutron star merger
  GW190425: Microphysics and mass ratio effects}.
\newblock {\em Mon. Not. Roy. Astron. Soc.} {\bf 2022}, {\em 516},~4760--4781.
\newblock {\url{https://doi.org/10.1093/mnras/stac2333}}.

\bibitem[Dietrich et~al.(2021)Dietrich, Hinderer, and
  Samajdar]{Dietrich:2020eud}
Dietrich, T.; Hinderer, T.; Samajdar, A.
\newblock {Interpreting Binary Neutron Star Mergers: Describing the Binary
  Neutron Star Dynamics, Modelling Gravitational Waveforms, and Analyzing
  Detections}.
\newblock {\em Gen. Rel. Grav.} {\bf 2021}, {\em 53},~27.
\newblock {\url{https://doi.org/10.1007/s10714-020-02751-6}}.

\bibitem[Agathos et~al.(2015)Agathos, Meidam, Del~Pozzo, Li, Tompitak, Veitch,
  Vitale, and Van Den~Broeck]{Agathos:2015uaa}
Agathos, M.; Meidam, J.; Del~Pozzo, W.; Li, T.G.F.; Tompitak, M.; Veitch, J.;
  Vitale, S.; Van Den~Broeck, C.
\newblock {Constraining the neutron star equation of state with gravitational
  wave signals from coalescing binary neutron stars}.
\newblock {\em Phys. Rev. D} {\bf 2015}, {\em 92},~023012.
\newblock {\url{https://doi.org/10.1103/PhysRevD.92.023012}}.

\bibitem[Bauswein et~al.(2013)Bauswein, Baumgarte, and
  Janka]{PhysRevLett.111.131101}
Bauswein, A.; Baumgarte, T.W.; Janka, H.T.
\newblock Prompt Merger Collapse and the Maximum Mass of Neutron Stars.
\newblock {\em Phys. Rev. Lett.} {\bf 2013}, {\em 111},~131101.
\newblock {\url{https://doi.org/10.1103/PhysRevLett.111.131101}}.

\bibitem[Oertel et~al.(2017)Oertel, Hempel, Kl\"ahn, and
  Typel]{RevModPhys.89.015007}
Oertel, M.; Hempel, M.; Kl\"ahn, T.; Typel, S.
\newblock Equations of state for supernovae and compact stars.
\newblock {\em Rev. Mod. Phys.} {\bf 2017}, {\em 89},~015007.
\newblock {\url{https://doi.org/10.1103/RevModPhys.89.015007}}.

\bibitem[Alford et~al.(2008)Alford, Schmitt, Rajagopal, and
  Sch\"afer]{RevModPhys.80.1455}
Alford, M.G.; Schmitt, A.; Rajagopal, K.; Sch\"afer, T.
\newblock Color superconductivity in dense quark matter.
\newblock {\em Rev. Mod. Phys.} {\bf 2008}, {\em 80},~1455--1515.
\newblock {\url{https://doi.org/10.1103/RevModPhys.80.1455}}.

\bibitem[{Noble} et~al.(2006){Noble}, {Gammie}, {McKinney}, and {Del
  Zanna}]{2006ApJ...641..626N}
{Noble}, S.C.; {Gammie}, C.F.; {McKinney}, J.C.; {Del Zanna}, L.
\newblock {Primitive Variable Solvers for Conservative General Relativistic
  Magnetohydrodynamics}.
\newblock {\em  Astrophys. J.} {\bf 2006}, {\em 641},~626--637.
\newblock {\url{https://doi.org/10.1086/500349}}.

\bibitem[Faber and Rasio(2012)]{faber2012binary}
Faber, J.A.; Rasio, F.A.
\newblock Binary neutron star mergers.
\newblock {\em Living Rev. Relativ.} {\bf 2012}, {\em 15},~1--83.

\bibitem[{Duez} et~al.(2005){Duez}, {Liu}, {Shapiro}, and
  {Stephens}]{2005PhRvD..72b4028D}
{Duez}, M.D.; {Liu}, Y.T.; {Shapiro}, S.L.; {Stephens}, B.C.
\newblock {Relativistic magnetohydrodynamics in dynamical spacetimes: Numerical
  methods and tests}.
\newblock {\em Phys. Rev. D} {\bf 2005}, {\em 72},~024028.
\newblock {\url{https://doi.org/10.1103/PhysRevD.72.024028}}.

\bibitem[{Font}(2000)]{2000LRR.....3....2F}
{Font}, J.A.
\newblock {Numerical Hydrodynamics in General Relativity}.
\newblock {\em Living Rev. Relativ.} {\bf 2000}, {\em 3},~2,
 \newblock {\url{https://doi.org/10.12942/lrr-2000-2}}.

\bibitem[Chang and Etienne(2020)]{Chang:2020ktl}
Chang, P.; Etienne, Z.
\newblock {General relativistic hydrodynamics on a moving-mesh I: Static
  space\textendash{}times}.
\newblock {\em Mon. Not. Roy. Astron. Soc.} {\bf 2020}, {\em 496},~206--214.
\newblock {\url{https://doi.org/10.1093/mnras/staa1532}}.

\bibitem[Kalinani et~al.(2025)]{Kalinani:2024rbk}
Kalinani, J.V.; Ji, L.; Ennoggi, L.; Lopez Armengol, F.G.; Sanches, L.T.; Tsao, B.-J.; Brandt, S.R.; Campanelli, M.;Ciolfi, R.; Giacomazzo, B.
\newblock {AsterX: A new open-source GPU-accelerated GRMHD code for dynamical
  spacetimes}.
\newblock {\em Class. Quant. Grav.} {\bf 2025}, {\em 42},~025016.
\newblock {\url{https://doi.org/10.1088/1361-6382/ad9c11}}.

\bibitem[Zhu et~al.(2024)Zhu, Fields, Zappa, Radice, Stone, Rashti, Cook,
  Bernuzzi, and Daszuta]{Zhu:2024utz}
Zhu, H.; Fields, J.; Zappa, F.; Radice, D.; Stone, J.; Rashti, A.; Cook, W.;
  Bernuzzi, S.; Daszuta, B.
\newblock {Performance-Portable Numerical Relativity with AthenaK}. \emph{arXiv} {\bf 2024}, arXiv:2409.10383.


\bibitem[Liebling et~al.(2020)Liebling, Palenzuela, and
  Lehner]{Liebling:2020jlq}
Liebling, S.L.; Palenzuela, C.; Lehner, L.
\newblock {Toward fidelity and scalability in non-vacuum mergers}.
\newblock {\em Class. Quant. Grav.} {\bf 2020}, {\em 37},~135006.
\newblock {\url{https://doi.org/10.1088/1361-6382/ab8fcd}}.

\bibitem[Dieselhorst et~al.(2021)Dieselhorst, Cook, Bernuzzi, and
  Radice]{dieselhorst21}
Dieselhorst, T.; Cook, W.; Bernuzzi, S.; Radice, D.
\newblock Machine {{Learning}} for {{Conservative-to-Primitive}} in
  {{Relativistic Hydrodynamics}}.
\newblock {\em Symmetry} {\bf 2021}, {\em 13},~2157.
\newblock {\url{https://doi.org/10.3390/sym13112157}}.

\bibitem[Ansel et~al.(2024)Ansel, Yang, He, Gimelshein, Jain, Voznesensky,
  et~al.]{ansel2024pytorch}
Ansel, J.; Yang, E.; He, H.; Gimelshein, N.; Jain, A.; Voznesensky, M.; Bao, B.; Bell, P.; Berard, D.; Burovski, E.; et~al.
\newblock PyTorch 2: Faster Machine Learning Through Dynamic Python Bytecode
  Transformation and Graph Compilation.
\newblock In Proceedings of the 29th ACM International
  Conference on Architectural Support for Programming Languages and Operating
  Systems (ASPLOS '24), ACM, {La Jolla CA USA, 27 April--1 May 2024.} %MDPI: We added the location and date of the conference. Please confirm.
  %Author: We confirm.
\newblock {\url{https://doi.org/10.1145/3620665.3640366}}.

\bibitem[Kastaun et~al.(2021)Kastaun, Kalinani, and Ciolfi]{Kastaun2021}
Kastaun, W.; Kalinani, J.V.; Ciolfi, R.
\newblock Robust {{Recovery}} of {{Primitive Variables}} in {{Relativistic
  Ideal Magnetohydrodynamics}}.
\newblock {\em Phys. Rev. D} {\bf 2021}, {\em 103},~023018.
\newblock {\url{https://doi.org/10.1103/PhysRevD.103.023018}}.

\bibitem[Banyuls et~al.(1997)Banyuls, Font, Ib{\'a}{\~n}ez, Mart{\'\i}, and
  Miralles]{Valencia1997}
Banyuls, F.; Font, J.A.; Ib{\'a}{\~n}ez, J.M.; Mart{\'\i}, J.M.; Miralles, J.A.
\newblock Numerical {3 + 1} General Relativistic Hydrodynamics: A Local
  Characteristic Approach.
\newblock {\em  Astrophys. J.} {\bf 1997}, {\em 476},~221.
\newblock {\url{https://doi.org/10.1086/303604}}.

\bibitem[Mart{\'i} and M{\"u}ller(2003)]{marti2003}
Mart{\'i}, J.M.; M{\"u}ller, E.
\newblock Numerical {{Hydrodynamics}} in {{Special Relativity}}.
\newblock {\em Living Rev. Relativ.} {\bf 2003}, {\em 6},~7.
\newblock {\url{https://doi.org/10.12942/lrr-2003-7}}.

\bibitem[Font(2008)]{font2008}
Font, J.A.
\newblock Numerical {{Hydrodynamics}} and {{Magnetohydrodynamics}} in {{General
  Relativity}}.
\newblock {\em Living Rev. Relativ.} {\bf 2008}, {\em 11},~7.
\newblock {\url{https://doi.org/10.12942/lrr-2008-7}}.

\bibitem[Janka et~al.(1993)Janka, Zwerger, and Moenchmeyer]{janka1993}
Janka, H.T.; Zwerger, T.; Moenchmeyer, R.
\newblock Does artificial viscosity destroy prompt type-II supernova
  explosions?
\newblock {\em Astron. Astrophys.} {\bf 1993}, {\em 268},~360--368.

\bibitem[Read et~al.(2009)Read, Lackey, Owen, and Friedman]{read2009}
Read, J.S.; Lackey, B.D.; Owen, B.J.; Friedman, J.L.
\newblock Constraints on a Phenomenologically Parametrized Neutron-Star
  Equation of State.
\newblock {\em Phys. Rev. D} {\bf 2009}, {\em 79},~124032.
\newblock {\url{https://doi.org/10.1103/PhysRevD.79.124032}}.

\bibitem[Schneider et~al.(2017)Schneider, Roberts, and Ott]{SRO2017}
Schneider, A.S.; Roberts, L.F.; Ott, C.D.
\newblock Open-Source Nuclear Equation of State Framework Based on the
  Liquid-Drop Model with {{Skyrme}} Interaction.
\newblock {\em Phys. Rev. C} {\bf 2017}, {\em 96},~065802.
\newblock {\url{https://doi.org/10.1103/PhysRevC.96.065802}}.

\bibitem[Wouters(2024)]{Wouters24}
Wouters, T.
\newblock {Machine} %MDPI: Unable to find relevant information for this reference. Please confirm if it is a reference that has been retracted. The withdrawn references need to be deleted. Please check.
%Author: As this is a Master’s thesis, we think it cannot be easily withdrawn or retracted. We do not have knowledge of its current official location on the web; however, a copy seems to be available on the author’s personal website: https://thibeauwouters.github.io/files/pdf/ML_algorithms_for_C2P_in_relativistic_hydrodynamics.pdf 
 Learning Algorithms for the Conservative-to-Primitive
  Conversion in Relativistic Hydrodynamics.
\newblock Master’s Thesis, KU Leuven, {Leuven, Belgium,} %MDPI: Newly added location information. Please confirm.
%Author: We confirm.
 2024.

\bibitem[Bernuzzi et~al.(2020)Bernuzzi, Breschi, Daszuta, Endrizzi, Logoteta,
  Nedora, Perego, Schianchi, Radice, Zappa, Bombaci, and Ortiz]{bernuzzi2020}
Bernuzzi, S.; Breschi, M.; Daszuta, B.; Endrizzi, A.; Logoteta, D.; Nedora, V.;
  Perego, A.; Schianchi, F.; Radice, D.; Zappa, F.;  et~al.
\newblock Accretion-Induced Prompt Black Hole Formation in Asymmetric Neutron
  Star Mergers, Dynamical Ejecta and Kilonova Signals.
\newblock {\em Mon. Not. R. Astron. Soc.} {\bf 2020},
  {\em 497},~1488--1507.
\newblock {\url{https://doi.org/10.1093/mnras/staa1860}}.

% \bibitem[{Semih Kacmaz}(2024)]{kacmaz2023c2pnet}
% \hl{Kacmaz, S.} %MDPI: Unable to find relevant information for this reference. Please confirm if it is a reference that has been retracted. The withdrawn references need to be deleted. Please check.
%Author: This was supposed to be a link to our codebase. We removed it from the references and added a direct link at the end of the Conclusions section (highlighted)
%   Recovery in Numerical Relativity. 2024.} %MDPI: We are sorry but we could not find the required information about this entry. Please provide more information about this article, whether it is a book (please provide the name and location of the publisher); online resource (please provide the URL of the website and the date it was accessed (Date Month Year)); or journal article (please provide the name of the journal, the year and volume in which it was published, and the page number). Please refer to https://www.mdpi.com/authors/references for full reference formatting guides.
  %Author: This was supposed to be a link to our codebase. We removed it from the references and added a direct link at the end of the Conclusions section (highlighted)


\bibitem[Boerner et~al.(2023)Boerner, Deems, Furlani, Knuth, and
  Towns]{access-ci}
Boerner, T.J.; Deems, S.; Furlani, T.R.; Knuth, S.L.; Towns, J. %MDPI: Please cite the reference in the text.
%Author: This omission was inadvertently introduced in the first proof. Apologies. Fixed. Now it is cited in the text.
\newblock ACCESS: Advancing Innovation: NSF’s Advanced Cyberinfrastructure
  Coordination Ecosystem: Services \& Support.
\newblock In Proceedings of the Practice and Experience in Advanced Research
  Computing 2023: Computing for the Common Good, New York, NY, USA, {23--27 July} 2023;
pp. 173–176.
\newblock {\url{https://doi.org/10.1145/3569951.3597559}}.

\bibitem[Hunter(2007)]{Hunter:2007}
Hunter, J.D.
\newblock Matplotlib: A 2D graphics environment.
\newblock {\em Comput. Sci. Eng.} {\bf 2007}, {\em
  9},~90--95.
\newblock {\url{https://doi.org/10.1109/MCSE.2007.55}}.

\bibitem[Waskom(2021)]{Waskom2021}
Waskom, M.L.
\newblock seaborn: Statistical data visualization.
\newblock {\em J. Open Source Softw.} {\bf 2021}, {\em 6},~3021.
\newblock {\url{https://doi.org/10.21105/joss.03021}}.

\end{thebibliography}

\end{document}